\newif\ifsingle

\newif\ifFullVersion

\ifsingle
\documentclass[11pt,draftclsnofoot, onecolumn]{IEEEtran}		
\else		
\documentclass[9pt,final, twocolumn]{IEEEtran}
\fi

\usepackage{times}
\usepackage{amsmath,dsfont}
\usepackage{amssymb,amsthm}
\usepackage{epsfig,verbatim}
\usepackage{subfigure}
\usepackage{setspace}
\usepackage{color}
\usepackage{cite}
\usepackage{epstopdf}
\usepackage{graphics}
\usepackage{accents}
\usepackage{acronym}
\usepackage[bookmarks,colorlinks]{hyperref}
\usepackage{booktabs}
\usepackage{mathtools}
\usepackage{algorithm}
\usepackage{algorithmic}
\usepackage{enumitem}

\newtheorem{lemma}{Lemma}

\ifsingle

\else

\fi 

\acrodef{adc}[ADC]{analog-to-digital convertor}
\acrodef{aoa}[AOA]{angle of arrival}
\acrodef{coa}[COA]{curvature of arrival}
\acrodef{crb}[CRB]{Cramér Rao Bound}
\acrodef{cs}[CS]{compressed sensing}
\acrodef{csi}[CSI]{channel state information}
\acrodef{dma}[DMA]{dynamic metasurface antenna}
\acrodef{dtft}[DTFT]{discrete-time Fourier transform}
\acrodef{dnn}[DNN]{deep neural network} 
\acrodef{gps}[GPS]{global positioning system} 
\acrodef{map}[MAP]{maximum a-posteriori probability}
\acrodef{snr}[SNR]{signal-to-noise ratio}
\acrodef{sinr}[SINR]{signal-to-interference-and-noise ratio}
\acrodef{bs}[BS]{base station} 
\acrodef{em}[EM]{electromagnetic} 
\acrodef{iot}[IOT]{Interent of Things}
\acrodef{mimo}[MIMO]{multiple-input multiple-output}
\acrodef{mse}[MSE]{mean-squared error}
\acrodef{pdf}[PDF]{probability density function}
\acrodef{rv}[RV]{random variable}
\acrodef{fec}[FEC]{forward error correction}
\acrodef{lti}[LTI]{linear time-invariant}
\acrodef{mle}[MLE]{maximum likelihood estimation}
\acrodef{wss}[WSS]{wide-sense stationary}
\acrodef{psd}[PSD]{power spectral density}
\acrodef{ris}[RIS]{reconfigurable intelligent surface}
\acrodef{ser}[SER]{symbol error rate} 
\acrodef{ber}[BER]{bit error rate} 
\acrodef{sgd}[SGD]{stochastic gradient descent} 
\acrodef{toa}[TOA]{time of arrival}
\acrodef{isi}[ISI]{intersymbol interference}  
\acrodef{awgn}[AWGN]{additive white Gaussian noise} 
\acrodef{ut}[UT]{user terminal} 
\acrodef{mmw}[mmWave]{millimeter wave}
\acrodef{6g}[6G]{sixth generation}

\IEEEoverridecommandlockouts

\title{Near-field Localization with Dynamic Metasurface Antennas
}
\author{  
	\IEEEauthorblockN{Qianyu Yang, Anna Guerra, Francesco Guidi, Nir Shlezinger, Haiyang Zhang, \\ Davide Dardari,
Baoyun Wang, and Yonina C. Eldar \\
	} 
	\thanks{

		}

	\vspace{-1.0cm}
	
}
\vspace{-0.75cm}

\begin{document}
	
	\maketitle
	\pagestyle{empty}
	\thispagestyle{empty}
\begin{abstract}
Sixth generation (6G) cellular communications are expected to support enhanced wireless localization capabilities. The widespread deployment of large arrays and high-frequency bandwidths give rise to new considerations for localization applications. First, emerging antenna architectures, such as dynamic metasurface antennas (DMAs), are expected to be frequently utilized thanks to the achievable high angular resolution and low hardware complexity.
Further, wireless localization is likely to take place in the radiating near-field (Fresnel) region, which provides new degrees of freedom, because of the adoption of arrays with large apertures.
While  current studies mostly focus on the use of costly fully-digital antenna arrays, in this paper we investigate how DMAs can be applied for near-field localization  of a single user. We use a direct positioning estimation method based on curvature-of-arrival of the impinging wavefront to obtain the user location, and characterize the effects of DMA tuning on the estimation accuracy. Next, we propose an algorithm for configuring the DMA to optimize near-field localization, by first tuning the adjustable DMA coefficients to minimize the estimation error using postulated knowledge of the actual user position. Finally, we propose a sub-optimal iterative algorithm that does not rely on such knowledge. 
Simulation results show that the DMA-based near-field localization accuracy could approach that of fully-digital arrays at lower cost.

\end{abstract}

\acresetall


\section{Introduction}

Radio positioning applications are expected to be notably enhanced and widely used in  \ac{6g}  networks~\cite{bibtex1,bibtex22,wang2022location}. The expected deployment of large antenna arrays and the usage of high frequencies will facilitate accurate radio
frequency (RF) positioning, especially in \ac{gps} denied scenarios. 
But these also give rise to two byproducts which have a dominant effect on the ability to carry out RF localization. First, the combination of large antennas and high frequencies implies that RF signaling likely takes place in the radiating near-field region, where the planar wavefront approximation commonly adopted in far-field systems does not hold~\cite{nepa2017near,zhang20226g}. Also, implementing arrays with a massive amount of elements is costly using conventional fully-digital designs, where each antenna is connected to a dedicated RF chain~\cite{ahmed2018survey}. Therefore, there is a need to understand the near-field positioning problem using antenna arrays that combine hybrid analog and digital  architectures. 

Source localization is typically based on two-step approaches, entailing a joint estimation of the \ac{aoa} and \ac{toa} \cite{bibtex3, bibtex4, bibtex19}. This requires precise synchronization and/or multiple access points participation \cite{bibtex5}, and typically tends to achieve sub-optimal performance compared to directly localization \cite{bibtex6}.
When moving from the far field to the radiating near field, the  new degrees of freedom, encapsulated in the spherical wavefront, can be exploited to enhance wireless localization, enabling  holographic localization \cite{bibtex2,bibtex23,DarDecGueGui:J22}. 

Near-field signal processing algorithms are typically designed by asusming that the received signal has a spherical wave rather than a plane wave  \cite{bibtex8}. A byproduct of this wave shapping is that it enables \ac{coa}-based localization, which is widely used in acoustic or microwave signalling \cite{ bibtex9, bibtex10}. \ac{coa}-based localization has only recently been considered for wireless communications at high-frequency bands \cite{bibtex16,bibtex17, bibtex11,bibtex12,bibtex18, bibtex13}. The recent studies of COA-based near-field localization for 6G include the characterization of the \acl{crb} \cite{bibtex16,bibtex17}; the investigation of \acl{ris}-assisted configurations \cite{ bibtex11,bibtex12,bibtex18}; and the derivations of  localization and tracking approaches in dynamic settings \cite{bibtex13}.

These aforementioned works on wireless localization all rely on fully-digital antenna architectures, which, while providing enhanced beamforming capabilities, may be too costly and power consuming for their employment in some 6G settings~\cite{zirtiloglu2022power}.
An emerging antenna technology that inherently operates with reduced RF chains without dedicated analog circuitry is based on \acp{dma} \cite{bibtex20}, realizing massive antenna arrays at reduced cost and with flexible analog beamforming and signal processing capabilities, which were shown to support high-rate communications with reduced RF chains in both conventional far-field \cite{shlezinger2019dynamic} and in near-field wireless communications \cite{bibtex15}. Yet, the usage of \acp{dma} for localization in \ac{6g} has not been studied to date, and is the focus of the current work. 

Here, we study location estimation for \ac{6g} systems using \acp{dma}. We propose an ad-hoc direct method based on \ac{coa} with \acp{dma} tuning to estimate the position of the user. Differently from common schemes using fully-digital antennas, we focus on localization with less complex \ac{dma}s that reduce RF chains, and illustrate how its dimensionality reduction of the received signal and inherent analog processing affect the \ac{mle} of the user location. Then, we formulate the analog processing of \acp{dma} for near-field localization as a form of precoding for near-field beam focusing~\cite{bibtex15}, which makes it possible to improve the received signal quality and to optimize the \ac{dma} coefficients setting for positioning.

The above \ac{dma} setting depends on the actual user position, which is unknown and must be estimated. Thus, we propose an iterative optimization algorithm for simultaneous localization and \ac{dma} tuning for the joint goal of accurate location estimation. 
We numerically demonstrate that the proposed solution can approach the performance achieved with costly fully-digital antennas by setting a suitable number of iterations that gradually refines the focused beam generated by the \ac{dma}. 

The rest of this work is organized as follows: Section~\ref{sec:Model} describes  the system and signal models and formulates the problem; the proposed \ac{dma} tuning algorithm is derived and evaluated in Sections~\ref{sec:Solution}-\ref{sec:Sims}, respectively. Finally, Section~\ref{sec:Conclusions} provides concluding remarks.



\section{System Model}
\label{sec:Model}

We consider a single user localization scenario in which a multi-antenna base station receives a pilot signal from the user and estimates its location. 
The base station is equipped with a DMA containing $ N $ receiving elements. The elements are placed along one-dimensional microstrips arranged in rows, each containing $N_{\rm e} $ elements with sub-wavelength space and connected to one RF chain output port. The number of microstrips is $N_{\rm d} ={ N }/N_{\rm e}$, i.e., the DMA only requires $N_{\rm d}$ RF chains. In contrast, in the common fully digital array which is composed of $N_{\rm d}$ antenna lines, each antenna element on the line with wavelength space is connected to a dedicated RF chain, thus the fully digital array of the same aperture as DMA is comprised of fewer antennas but requires more RF chains.

The array aperture and the signaling frequency are assumed to be such that the transmitting user resides in the radiating near-field, i.e., in the Fresnel region (which can be in distances of the order of tens and even hundreds of meters in some \ac{6g} settings \cite{zhang20226g}). 
Since the user is within the Fresnel region of the \ac{dma}, the received signal exhibits a spherical wavefront, as displayed in Fig.~\ref{fig2}. 

\subsection{Received Signal Model} \label{sub:DMA}

We suppose a narrowband pilot signal at frequency $f_{\rm p}$ is emitted by the source with power normalized to 1. The signal received by the $l$-th antenna of the $i$-th line microstrip at the generic discrete time instant is given by
\begin{align}
{ x}_{i, l}={\rm g}_{i,l} x_0 +z_{i,l}
\label{receive}
\end{align}
where ${\rm g}_{i, l} = a_{i, l}  e^{-j v_{i, l}}$ represents the channel component, with ${ a}_{i, l}$ and $v_{i, l}$ denoting the channel gain coefficient and the phase due to the distance traveled by the signal, respectively. $x_0$ is the source pilot signal, and $z_{i,l}$ is an additive thermal noise with power $\delta^2$. The phase is given by $v_{i, l} \triangleq  2 \pi  f_{\rm p} \frac{ d_{i, l}}{c}$, where $ d_{i, l}$ is the distance between the corresponding antenna and the source, and $c$ is the speed of light. 
In vector form, \eqref{receive} becomes 
\begin{equation} 
\label{x}
{\bf x}={\bf g} x_0 + {\bf z}, 
\end{equation}
where $ {\bf g}=\left[ {\rm g}_{1,1}, \cdots,  {\rm g}_{i,l}, \cdots, {\rm g}_{N_{\rm d},N_{\rm e}} \right] \in \mathbb{C}^{ N}$ and $ {\bf z}=\left[z_{1,1}, \cdots, z_{i,l}, \cdots, z_{N_{\rm d},N_{\rm e}} \right] \in \mathbb{C}^{N}$ denote the channel vector and noise vector, respectively.

\begin{figure}
\centering
\includegraphics[scale=0.6]{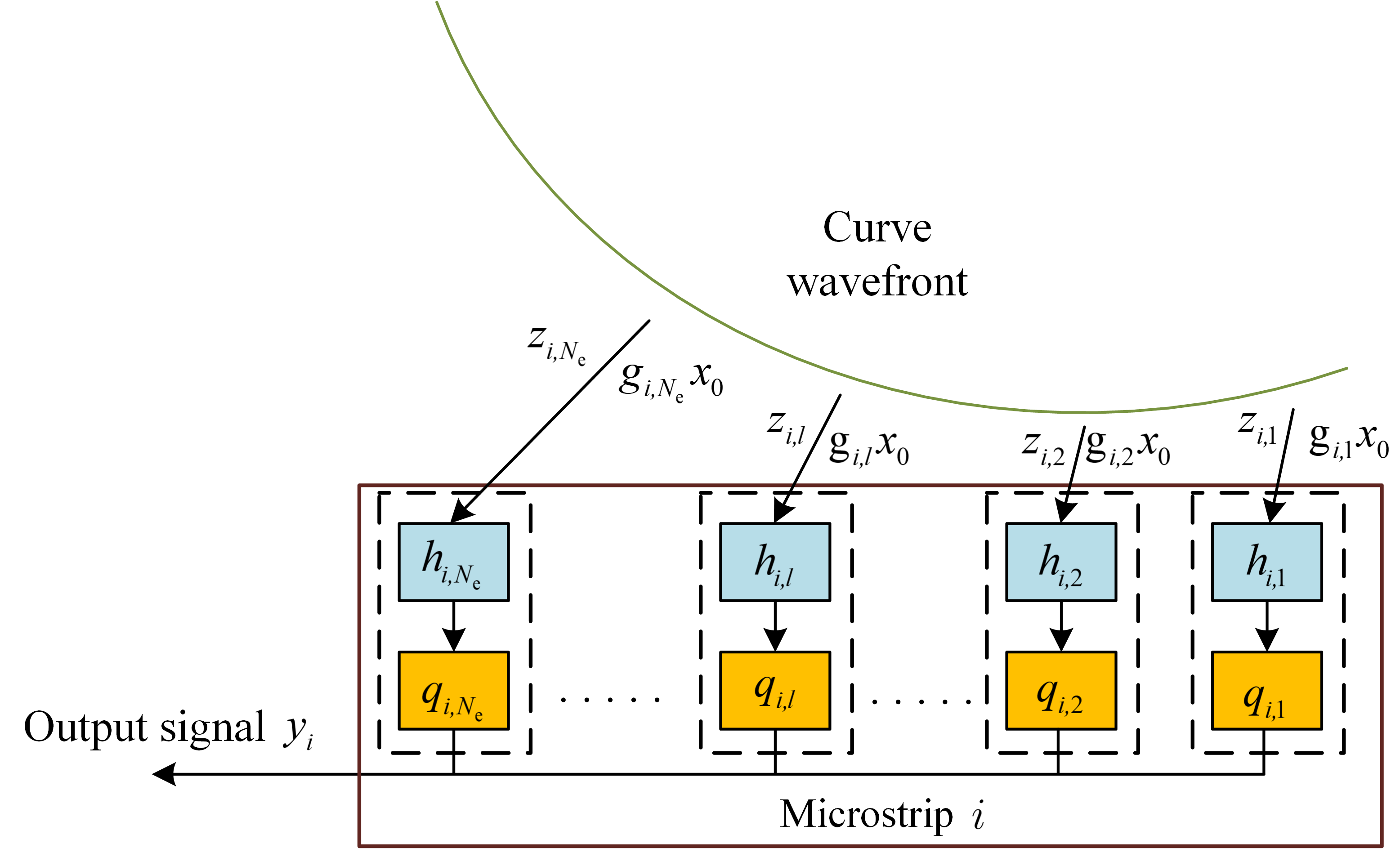}
\caption{Signal reception at the $i$-th microstrip} 
\label{fig2}
\end{figure}

The signals impinging on the \ac{dma} elements propagate inside the waveguide and are captured at the output port. The output of each microstrip is thus modelled as a weighted sum of these received signals. Hence, the signal captured at the RF chains after undergoing the elements response and propagating inside the waveguide is given by:
\begin{equation} 
\label{y}
{\bf y}={\bf Q} {\bf H}{\bf x}, 
\end{equation}
where $\bf H$ is a $ N \times N$ diagonal matrix with diagonal elements ${\bf H}_{\left((i-1) {N_{\rm e}}+l,(i-1) {N_{\rm e}}+l\right)}={ h}_{i, l}$, with ${ h}_{i, l}$ encapsulating the effect of signal propagation inside the microstrip, and is modelled as \cite{shlezinger2019dynamic}
\begin{align}
\label{attenuation}
&{ h}_{i, l}=e^{-\rho_{i, l}\left(\alpha_{i}+j \beta_{i}\right)},
\end{align}
where $\alpha_{i}$ is the waveguide attenuation coefficient, $\beta_{i}$ is the wavenumber, and $\rho_{i, l}$ denotes the distance of the $l$-th element from the output port of the $i$-th microstrip.
The matrix ${\bf Q} \in \mathbb{C}^{N_{\rm d} \times N}$  in \eqref{y} denotes the configurable weights of the \acp{dma}. Since we focus on narrowband signalling, for which the elements can be approximated with the Lorentzian-constrained form \cite{bibtex15}, ${\bf Q}$ obeys the following structure:
\begin{equation}
\label{stru}
{\bf Q}_{n, (i-1) {N_{\rm e}}+l}= \begin{cases}{ q}_{i, l} \in \mathcal{Q} & i=n, \\ 0 & i \neq n,\end{cases}
\end{equation}
where ${ q}_{i, l}$ denotes the tunable response of corresponding antenna, and its feasible settings are
\begin{align}
\label{lim}
&{ q}_{i, l} \in \mathcal{Q} \triangleq\left\{\frac{j+e^{j \phi_{i,l}}}{2} \mid \phi_{i,l} \in[0,2 \pi]\right\}, &&\forall i, l.
\end{align}
The received signal model is depicted in the lower part of
Fig.~\ref{fig2}.

\subsection{COA-Based Localization} \label{sub:model}
\begin{figure}
\centering
\includegraphics[scale=0.8]{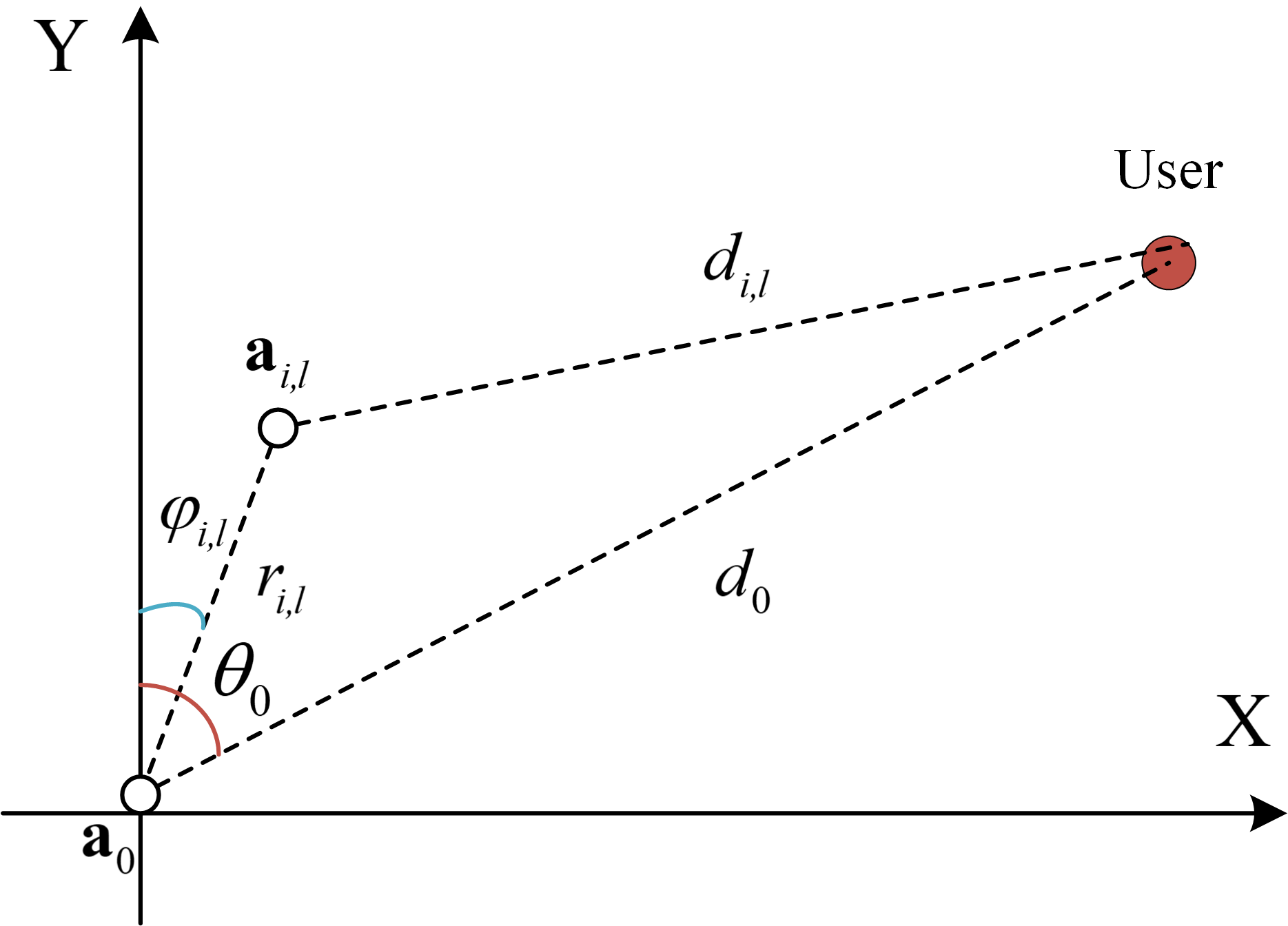}
\caption{Relative position relationships between source and array. The $\mathbf{a}_0$ is reference antenna, whereas the antenna of the $i$-line and $l$-row is at $\mathbf{a}_{i,l}$. 
}
\label{fig3}
\end{figure}

We consider the task of estimating the user position based on \eqref{y}, while tuning the DMA weights ${\bf Q}$ to achieve an accurate estimate. 
The fact that the communication takes place in the radiating near-field can be exploited to facilitate localization based on the COA of the impinging signal at different elements. To see this, consider a 2D case as illustrated in Fig.~\ref{fig3}. We set a reference point (i.e., antenna) for the receiving array, which is located at distance ${d}_0$, forming an angle $\theta_0$ with respect to the source (see Fig.~\ref{fig3}). Further, we define $\left( {r}_{i,l} ,\varphi_{i,l}\right)$ to indicate the position of $l$-th antenna in the $i$-th microstrip with respect to the array center. According to the triangular relationship between reference points, the antenna and the user, ${ d}_{i, l}$ could be confirmed by $d_0$ and $\theta_0$ as \cite{ bibtex13}:
\begin{equation}
\label{loca1}
{ d}_{i, l}\left({ d}_0 ,\theta_0 \right)=\sqrt{{ r}_{i, l }^{2}+{ d}_{0}^{2}-2 { r}_{i, l } { d}_{0} \cos \left(\varphi_{i, l }-\theta_{0}\right)},
\end{equation}
the operation in the radiating near-field implies that none of the terms in \eqref{loca1} can be considered as being negligible.

Equation \eqref{loca1} determines the relationship between the user location and the phase profile, which is highly nonlinear, and is retained at the output of the DMA by \eqref{y}. Therefore, a possible solution for estimating the user position is to compute the MLE, given by
\begin{equation}
\label{que1}
\left(d^{\star}, \theta^{\star} \right)=\underset{\left({ d} ,\theta\right)}{\arg \max } \log p\left({\bf y} ; { d} ,\theta \right),
\end{equation}
where $ p\left({\bf y} ; { d} ,\theta \right)$ is the likelihood function of $\bf y$, $\left(d^{\star}, \theta^{\star} \right)$ and $\left(d, \theta \right)$ are the estimated and the possible user positions in polar coordinates. 

For the fully digital array case, each antenna is connected to a dedicated RF chain (without analog combining operation), leading to the received signal ${\bf y}={\bf x}$. In this case, the likelihood function in \eqref{que1} is of $\bf x$, which can be expressed as \cite{ bibtex9}:
\begin{equation}
\label{logp-f}
\log p\left({\bf x} ; d ,\theta \right) \propto \sum \Vert {\bf P}_{\rm FD}{\left( d ,\theta \right)} {\bf x} \Vert^{2},
\end{equation} 
where ${\bf P}_{\rm FD}{\left( d ,\theta \right)}={\bf s} \left[ {\bf s}^{H} {\bf s} \right]^{-1} {\bf s}^{H}$ is the projection operator of ${\bf s }$, with ${\bf s }= \left[ {\rm a}_{1, 1} e^{-j v_{1,1}(d,\theta) }, \cdots, {\rm a}_{N_{\rm d},N_{\rm e}} e^{-j v_{N_{\rm d},N_{\rm e}}(d,\theta) } \right]^T$ denoting the steering vector of the array (in near field, the steering vector concept loses its usual meaning as it cannot be identified an unique steering direction. Despite that, we still adopt this term in a wide sense). Note that $\bf s =g$ only if $d=d_0$ and $\theta=\theta_0$. 

Likewise, for the DMA case, according to \eqref{y}, the log-likelihood function of ${\bf y}$ can be expressed as:
\begin{equation}
\label{logp-d}
\log p\left({\bf y} ; d, \theta \right) \propto \sum \Vert {\bf P}_{\rm DMA}{\left( d, \theta, {\bf Q} \right)} {\bf y} \Vert^{2},
\end{equation}
where ${\bf P}_{\rm DMA}{\left( d ,\theta, {\bf Q} \right)} = {\bf Q}{\bf H}{\bf s} \left[ {\bf s}^{H}{\bf H}^H{\bf Q}^{H} {\bf Q}{\bf H}{\bf s} \right]^{-1} {\bf s}^{H}{\bf H}^H{\bf Q}^{H}$. Compared to \eqref{logp-f}, ${\bf P}_{\rm DMA}{\left( d ,\theta, {\bf Q} \right)}$ is additionally controlled by $\bf Q$.

Using \eqref{logp-d}, one can compute \eqref{que1} and localize the user when ${\bf Q}$ is fixed, i.e., its computation requires a given ${\bf Q}$. Therefore, the fact that the accuracy of this estimate depends on ${\bf Q}$, allows us to use the above MLE as guidelines for tuning the DMA along with the localization task, as detailed next.

\section{Joint Localization and DMA Tuning}
\label{sec:Solution}
	
\subsection{DMA Tuning} \label{sec: single-user}
We first consider only the DMA tuning to optimize the MLE~\eqref{que1}, ignoring for now the fact that its computation requires knowledge of the user position.
For ease of analysis, as in \cite{ bibtex9}, we rewrite \eqref{logp-d} as
\begin{equation}
\label{loglike}
\log p\left({\bf y} ; d ,\theta \right) \propto \operatorname{tr} \left [ {\bf P}_{\rm DMA}\left(d, \theta, {\bf Q} \right) {\bf R}  \right ],
\end{equation}
where ${\bf R} = {\bf Q}{\bf H} \left( {\bf g}{\bf g}^H+ \delta^2 {\bf I}_{ N} \right){\bf H}^H {\bf Q}^H$ is the covariance of ${\bf y}$. For convenience, we define a function $f\left( d, \theta, {\bf Q}\right)$ as:
\begin{equation}
\label{fun_f}
f\left( d, \theta, {\bf Q}\right) = \operatorname{tr} \left [ {\bf P}_{\rm DMA}\left(d, \theta, {\bf Q} \right) {\bf R}  \right ].
\end{equation}

The MLE computation can be regarded as a search process over distance and angle to maximize \eqref{fun_f}, and the maximum value of \eqref{fun_f} is expected to converge to the true user position, i.e., $d^{\star}=d_0$ and $\theta^{\star}=\theta_0$. 
Therefore, we aim to design $\bf Q$ to maximize \eqref{fun_f} at the actual position, i.e., considering the following problem:
\begin{equation}
\label{max_pro}
\begin{aligned}
&\max _{\bf Q} ~f\left( d_0, \theta_0, {\bf Q}\right). 
\end{aligned}
\end{equation} 
Using ${\bf P}_{\rm DMA}\left(d_0, \theta_0, {\bf Q} \right) {\bf Q} {\bf H} {\bf g}{\bf g}^H {\bf H}^H {\bf Q}^H = {\bf Q} {\bf H} {\bf g}{\bf g}^H {\bf H}^H {\bf Q}^H$, the problem \eqref{max_pro} can be explicitly rewritten to
\begin{equation}
\label{max_prop2}
\begin{aligned}
\max _{\bf Q} ~&\operatorname{tr} \left [ {\bf Q} {\bf H} {\bf g}{\bf g}^H {\bf H}^H {\bf Q}^H \right] \\
&+\operatorname{tr} \left [ {\bf P}_{\rm DMA}\left(d_0, \theta_0, {\bf Q} \right) \delta^2 {\bf Q} {\bf H}{\bf H}^H {\bf Q}^H \right]. 
\end{aligned}
\end{equation}



The Lorentzian-constrained $\mathcal{Q}$ form in \eqref{lim} makes the problem \eqref{max_prop2} still difficult to solve. Following \cite{bibtex15}, we tackle this  by relaxing the Lorentzian constraint to the phase-only weights constraint with constant amplitude and arbitrary phase, i.e.,
\begin{align}
\label{limf0}
&{ q}_{i, l} \in \mathcal{F} \triangleq\left\{e^{j \phi_{i,l}} \mid \phi_{i,l} \in[0,2 \pi]\right\}. 
\end{align}
We then project the solution onto $\mathcal{Q}$ as the approximate solution. The relaxed problem is given by 
\begin{equation}
\label{max_prop}
\begin{aligned}
&\max _{\bf Q} ~\operatorname{tr} \left [ {\bf Q} {\bf H} {\bf g}{\bf g}^H {\bf H}^H {\bf Q}^H \right] \\ \text{s.t.}~~ &{\bf Q}_{n, (i-1) {N_{\rm e}}+l}= \begin{cases}{ q}_{i, l} \in \mathcal{F} & i=n. \\ 0 & i \neq n.\end{cases}
\end{aligned}
\end{equation}
The second factor of problem \eqref{max_prop2} is omitted in \eqref{max_prop} as it is a constant. This is because, ${\bf Q} {\bf H}{\bf H}^H {\bf Q}^H = \sum_{l=1}^{l= N_{\rm e}}{h_{i,l}^2}{\bf I}_{N_{\rm d}}$ and $\operatorname{tr} \left [ {\bf P}_{\rm DMA}\left(d_0, \theta_0, {\bf Q} \right)\right]=1$.
The solution to the relaxed problem \eqref{max_prop} is stated in the following:
\begin{lemma}
 \label{thm:B}
The solution to \eqref{max_prop}, denoted ${\bf Q}^*$, is obtained by setting each non-zero element to ${ q}_{i, l}^{*}=e^{j \psi_{i, l}^{*}}$, with $\psi_{i, l}^{*}=v_{i,l} + \rho_{i, l} \beta_{i}$ 
\end{lemma}
\begin{IEEEproof}
Rewrite \eqref{max_prop} to scalar form to remove structural constraints \eqref{stru}:
\begin{equation}
\label{max2}
\max _{{ q}_{i, l}} \sum_{i=1}^{N_{\rm d}}\left|\sum_{l=1}^{N_{\rm e}} { q}_{i, l} h_{i,l}  {\rm g}_{i, l} x_0 \right|^{2} , \quad \text { s.t. } {q}_{i, l} \in \mathcal{F}.
\end{equation}
That is, it is decomposed into $N_{\rm d}$ subproblems by \eqref{max2}. Substituting the expression of ${ g}_{i, l}$ in \eqref{receive} into \eqref{max2}, we have the $i$-th subproblem:
\begin{equation}
\label{max3}
\max _{\psi_{i, l}} \left|\sum_{l=1}^{N_{\rm e}}  x_0 e^{-\rho_{i, l}\alpha_{i}} e^{- j v_{i,l}} e^{-j\rho_{i, l} \beta_{i}} e^{j \psi_{i, l}} \right|^{2}. 
\end{equation}
Hence, according to the triangle inequality, the solution to \eqref{max3} is: 
\begin{equation}
\label{phase}
\psi_{i, l}^{*}=v_{i,l} + \rho_{i, l} \beta_{i}.
\end{equation}
This proves the lemma.
\end{IEEEproof}
As ${ q}_{i, l}^{*}$ in Lemma~\ref{thm:B} does not satisfy the Lorentzian form, we project it onto \eqref{lim} following \cite{bibtex15}. The resulting weight is ${ \hat q}_{i,l} = \frac{j+e^{j\psi_{i, l}^{*}}}{2}$.

\subsection{Alternating Localization and DMA Tuning} \label{sec:multi-user}
The DMA tuning detailed above requires knowledge of the user position, whose estimation is the core of the localization problem. However, it can be adopted for joint DMA tuning and localization via alternating optimization. The rationale here stems from the fact that  the design of  $\bf Q$ for \ac{coa}-based localization realizes a form of beamforming, which in the radiating near-field specializes in beam focusing \cite{nepa2017near}. Consequently, \eqref{max_pro} can be regarded as designing $\bf Q$ to focus the received beam towards a given location, and iteratively refine the specific location via the \ac{mle}.  
Therefore, for the \ac{mle} process in \eqref{loglike}, even when the focusing position at the array receiving end deviates from the actual source position, the estimated results are still improved, which means that, once we obtain a rough estimation of the source location, we can constantly approach the desired actual source position by updating the focus position.

The resulting alternating localization and DMA tuning algorithm is summarized as Algorithm~\ref{alg:Alternating}. 
In Algorithm~\ref{alg:Alternating}, we utilize the receiving vector from random coefficient response performing the initial position estimation, and sample new observations after updating the coefficients of $\bf Q$ with the results of the previous estimation. We can repeat the above steps at each receiving time slot so that the estimated result is close to the actual source location gradually.
\begin{algorithm}
\caption{Iterative optimization algorithm}
\label{alg:Alternating}
\begin{algorithmic}[1] 
\renewcommand{\algorithmicrequire}{\textbf{Initialize:}} 
\REQUIRE Position mark $\left( d ,\theta \right)$
\renewcommand{\algorithmicrequire}{\textbf{Ensure:}} 
\REQUIRE Set $k=0$, the number of iterations $ K$. Set the matrix ${\bf Q}^0 \in \mathbb{C}^{N_{\rm d} \times N}$ constrained by the array architecture with random coefficient, and obtain the initial output vector ${\bf y}^0 $.
\renewcommand{\algorithmicrequire}{\textbf{While} {$k \leq  K$} \textbf{do}} \REQUIRE  
    \STATE Estimate $\left( d^k ,\theta^k \right)$ by maximizing $f\left( d, \theta, {\bf Q}^k\right)$ (equivalently maximize the MLE defined in \eqref{que1}).
    \STATE Get ${\bf Q}^{k+1}$ by maximizing $f\left( d^k, \theta^k, {\bf Q}\right)$ through  solving \eqref{max_prop2}.
    \STATE Update output vector ${\bf y}^{k+1}$ from ${\bf Q}^{k+1}$ by \eqref{y}.
    \STATE $k=k+1$
\renewcommand{\algorithmicrequire}{\textbf{end while}} \REQUIRE
\STATE Estimate $\left( d^{\star}, \theta^{\star} \right)$ = $\left( d^{K} ,\theta^{ K} \right)$
\end{algorithmic}
\end{algorithm}




\section{Numerical Evaluations}
\label{sec:Sims}

	
In this section, we evaluate the localization performance under multiple array architectures for different SNR scenarios. We consider localizing a user in a 2D (X-Y) plane, while the antenna is placed on the on Y-Z plane (with each microstrip lying on the Z axis and different microstrips stacked along the Y axis). The carrier frequency is $f_{\rm p} = 28$ GHz, corresponding to a  signal wavelength of $\lambda = 0.01$ m.
The DMA array is comprised of $N_{\rm d } = 5$ microstrips, i.e., it only requires 5 RF chains, with each microstrip length $L = 24$ cm. For the antenna spacing, we consider both sub-wavelength $ \lambda /4 $ antenna spacing and $\lambda /2$ antenna spacing (as in the fully digital case). We compare the results with a fully digital array of the same dimensions as the DMA and  $\lambda /2 $ antenna spacing. The fully digital has $N_{\rm F} = 240$ elements (and hence 240 RF chains), while the DMA has $5$ RF chains supporting  $N=240$ elements (for  $\lambda /2$ spacing), while stacking $N=480$ elements in the same aperture for   $ \lambda /4 $ spacing.   

The source position denoted $\left( d_0, \theta_0 \right)$ is located in the Fresnel region of the array, i.e., $d_0$ is smaller than the Fraunhofer limit $d_{\rm F}$ which, under the current setting, equals $d_{\rm F} =24 $ meters. In the following study, we set $d_0 = 6$ meters and  $\theta_0 = \pi / 3$.

\begin{figure}[t!]
\centering 
\includegraphics[width=0.85\columnwidth]{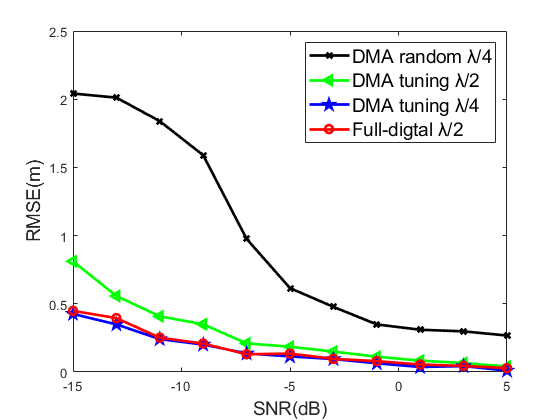}
\caption{Distortion contrast of multiple contrast schemes under different SNR with $d_0 = 0.25 \, d_{\rm F}$ meters, $d_{\rm F}=24$ meters, $\theta_0 = \pi /3$. 
}
\label{fig7}
\end{figure}

\begin{figure} 
\centering 
\subfigure[$d_0 = 0.25 \, d_{\rm F}$ meters, $d_{\rm F}=24$ meters with $L = 24$ cm]{ 
    \label{fig:subfig:near-field}
    \includegraphics[width=3in]{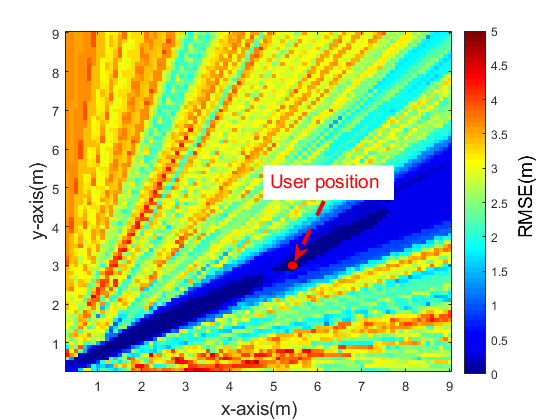} 
  } 
  \vspace{-0.2cm}
  \subfigure[$d_0 = 3 \, d_{\rm F}$ meters, $d_{\rm F}=2$ meters with $L = 5$ cm]{ 
    \label{fig:subfig:far-field} 
    \includegraphics[width=3in]{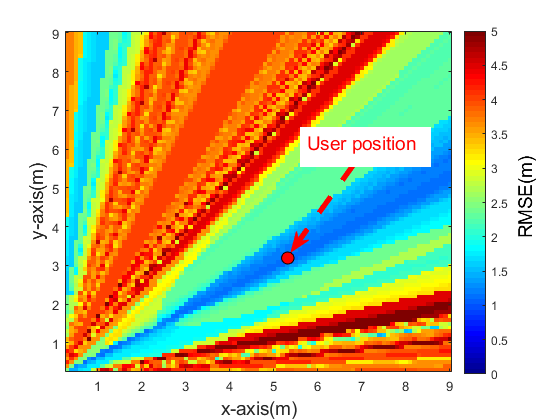} 
  }
 \caption{The heatmap for position estimate RMSE under different initial expected positions of Algorithm~\ref{alg:Alternating}: (a) $d_0 = 0.25 \, d_{\rm F}$ meters, $d_{\rm F}=24$ meters; (b) $d_0 = 3 \, d_{\rm F}$ meters, $d_{\rm F}=2$ meters. The different $d_{\rm F}$ in (b) is realized by reducing the array aperture. $\theta_0 = 60^\circ$, SNR is fixed as $-5$ dB. 
  } 
  \label{fig8} 
\end{figure}

We first show the effectiveness of the \ac{dma} tuning scheme we proposed in Fig. \ref{fig7} which reports the root mean square error (RMSE) of location estimation via different schemes under different SNRs. To indicate the performance improvement from tuning design, the DMA random scheme and the fully digital scheme are also provided for comparison, where the DMA random scheme is obtained by setting a set of random coefficients for $\bf Q$, while the fully digital scheme localizes the user with MLE method \cite{bibtex13} based on the fully digital array we just set.
As we can see, the DMA random scheme yields higher RMSE compared to the fully digital schemes, especially in the low SNR case. This is because the tunable coefficients matrix in the DMA random scheme represents only an interference effect for the MLE process. By tuning the DMA using Algorithm~\ref{alg:Alternating}, one significantly improves the RMSE, however, the RMSE of the DMA tuning scheme with $\lambda/2$ antenna spacing still higher than the fully digital schemes due to the reduction of signal dimension, while the DMA tuning scheme with $\lambda/4$ achieve the performance similar to that of the full-digital array scheme, in fact, in low SNRs the increased number of elements combined with their proper tuning via Algorithm~\ref{alg:Alternating} can achieve slightly better localization accuracy compared with fully-digital arrays utilizing $\times 48$ more RF chains. 

To indicate that the joint localization and DMA tuning realize a form of near-field beam focusing, we provide the RMSE heatmaps with the RMSE around the receiver, as shown in Fig.~\ref{fig8}. We consider a $9\times 9 ~ {\rm m}^2$ grid of points with resolution of $0.1$ meters, with the array located in the origin. Then, we set each grid point as the initial expected position of Algorithm~\ref{alg:Alternating} to estimate the actual user position. 
The user position remains at $\left( d_0, \theta_0 \right) = (6,\pi /3)$, and we consider two DMAs, with different apertures: one which aperture of  $24$ cm, for which $d_0 = 0.25 \, d_{\rm F}$, i.e., it operates in the radiating near-field; the other has a smaller aperture of $5$ cm, resulting in far-field operation as $d_0 = 3 \, d_{\rm F}$. The estimated RMSEs for each initial position of Algorithm~\ref{alg:Alternating} are depicted in Fig.~\ref{fig:subfig:near-field}  for the near-filed \ac{dma} and in Fig.~\ref{fig:subfig:far-field} for the far-field one. 
From Fig.~\ref{fig:subfig:near-field}, we can see that even if the initial expected position is not the actual user position, the  RMSE would be significantly improved once the initial position locates in the beam focusing area between the user and the array. Especially, the area around the user position obtains the lowest value of RMSE,  allowing to converge to the actual user position from different initial position of Algorithm~\ref{alg:Alternating}. In contrast, when the user is located in the far-field of the array as  in Fig.~\ref{fig:subfig:far-field}, the RMSE improvement deteriorates to along one direction only case, even if the initial position is the actual user position, it yields the same RMSE as any point in that direction. This is because the CoA reduces to the AOA in the far-field case, losing the ability to estimate the distance.

\section{Conclusions}
\label{sec:Conclusions}

In this work, we studied near-field source position estimation based on a DMA array. We presented a model for DMA-based radiating near-field position estimation systems, and utilized a direct estimation method based on COA for localization. We then formulated the optimization of the DMA tuning to improve the accuracy of the position estimate, and proposed an efficient algorithm for joint position estimate and DMA tuning. Numerical results demonstrated that the DMA tuning design could significantly improve the near-field localization performance reaching and overtaking that of a fully digital implementation with at least one order of RF chains reduction. 


	\bibliographystyle{IEEEtran}
	\bibliography{IEEEabrv,refs}

\end{document}